\documentclass[prd,
]
{revtex4}
\usepackage{graphicx}
\usepackage{dcolumn}
\usepackage{bm}
\usepackage{amssymb}
\usepackage{amsmath}
\usepackage{mathrsfs}
\usepackage{color}
\usepackage[toc,page]{appendix}
\newcommand{\bl}[1]{{\color{blue}#1}}
\newcommand{\ma}[1]{{\color{magenta}#1}}
\newcommand{\re}[1]{{\color{red}#1}}

\newcommand{\cy}[1]{{\color{cyan}#1}}
\DeclareMathOperator\arctanh{arctanh}

\begin{document}

\title{Hodograph  solutions of the  wave equation \\
of  nonlinear electrodynamics in  the quantum vacuum }
\author{Francesco Pegoraro}
\affiliation{Enrico Fermi Department of Physics, University of Pisa, Italy
and National Research Council, National Institute of Optics, via G. Moruzzi 1, Pisa, Italy}
\author{Sergei V. Bulanov}
\affiliation{Institute of Physics of the ASCR, ELI--Beamlines project, Na Slovance 2, 18221, Prague, Czech Republic}
\affiliation{National institutes for Quantum and Radiological Science and Technology (QST),
Kansai Photon Science Institute, 8--1--7 Umemidai, Kizugawa, Kyoto 619--0215, Japan}
\affiliation{Prokhorov  General Physics Institute of RAS, Vavilov Str.~38, Moscow 119991, Russia}
\date{\today}

\begin{abstract}

The process of photon-photon scattering in vacuum is investigated analytically  in the long-wavelength limit 
within the framework of the Euler-Heisenberg Lagrangian. 
In order to solve the nonlinear  partial differential equations  (PDEs)  obtained from 
this  Lagrangian 
 use is made of  the hodograph transformation. This transformation  makes it possible to turn a system of quasilinear PDEs 
into a  system of linear PDEs. 
\, \,  Exact solutions of the equations describing  the nonlinear interaction of electromagnetic waves in vacuum
in a one-dimensional configuration are obtained and analyzed.
 
\end{abstract}

\bigskip
\maketitle

\section{Introduction}
\label{sec:intro}

Perturbation theory has proven to be extremely successful in obtaining a number of prominent results in 
 quantum field theories (QFTs) \cite{QFT1, BLP-QED, QFT2, QFT3}. In spite of these achievements, as is well known, 
 perturbation theory is only  valid provided the interaction is weak and  thus  it cannot provide a full description
of a QFT \cite{FJD52, R70}. For this reason  the nonperturbative behavior of QFTs has attracted a great deal of attention for decades \cite{FS13}. As   examples of physical objects typical for QFTs and classical 
mechanics of continous media whose theoretical 
description cannot be obtained within the framework of pertutbation theory we may list the breaking of nonlinear waves,  solitons, instantons, etc. \cite{Whitham, VR02, NNR98, SS00, YuKPA03}. 

In quantum electrodynamics (QED)  perturbation theory breaks in the limit of strong electric fields, when  the electric field $E$ approaches  the  critical field of quantum electrodynamics~\cite{SAU31, HeisenbergEuler}
\begin{equation}
E_S=m_e^2c^3/e \hbar 
\label{eq:ES}
\end{equation}
and/or the photon energy becomes substantially large, i.e. 
for $\alpha \chi_{\gamma}^{2/3}\geq 1$~\cite{R70}   where 
$\alpha=e^2/\hbar c$ is the fine structure constant, $ \chi_{\gamma}=\hbar \sqrt{(F_{\mu \nu} k^{\mu})^2}/m_e cE_S$ 
is the so called nonlinear quantum parameter (see Refs.~\cite{BLP-QED, R70}), $F_{\mu \nu}$ is the electromagnetic  
field tensor, and $\hbar k^{\mu}$ is the four-momentum of the photon. The electrom mass and electric charge are 
$m_e$ and $e$, respectively, $c$ is the speed of light in vacuum, and $\hbar$ is the Planck constant.
The critical field  corresponds  to the electric field  that, acting on  the electron  charge $e$, would produce  a  work 
equal to the electron rest mass energy $m_ec^2$ 
over a distance equal to the Compton wavelength $ \lambdabar_C=\hbar/m_ec$. Here $\hbar$ is the 
reduced Planck constant, $e$ and $m_e$ are the electron electric charge and mass, and $c$ is the speed of light in 
vacuum (see for details Refs.~\cite{BLP-QED, SAU31, HeisenbergEuler, GVD09}).
The corresponding wavelength and  intensity of electromagnetic radiation are 
$\lambda_S=2 \pi \lambdabar_C\approx 2\times 10^{-10}$cm 
and $I_S=c E_S^2/4\pi\approx 10^{29}\,$W/cm$^2$, respectively.

One of the most remarkable effects predicted in QED is the vacuum polarization connected with light-light scattering 
and pair production from vacuum. In classical electrodynamics  electromagnetic waves do not interact in vacuum. 
On the contrary, in QED  photon-photon scattering 
can take place  in vacuum via the generation of virtual electron-positron pairs. This interaction gives rise to vacuum 
polarization and  birefringence,  to the Lamb shift,  to a modification  of the Coulomb field, 
and to many other phenomena  \cite{BLP-QED}. Photon-photon scattering was observed in collisions of heavy ions 
accelerated in standard  particle accelerators   (see review article \cite{Baur} 
and the results of the experiments obtained with the ATLAS detector at the Large Hadron Collider
 \cite{ATLASScattering}). 
 
Photon-photon interaction provides  a tool for the search for  new physics \cite{Baur, PLowdon}:
further studies  of this  process will make it possible to test extensions of the Standard Model in which new particles 
contribute to the interaction loop diagrams \cite{Inada}. Using the Euler-Heisenberg Lagrangian~
\cite{HeisenbergEuler, Schw51}, which describes the vacuum polarization and electron-positron pair generation by 
super-strong electromagnetic field in vacuum~\cite{ GVD09,  Borel}  also provides one of the most developed 
approaches for  studying non-perturbative processes in QFT, when finding exact solutions of nonlinear problems 
cannot be underestimated.

The increasing  availability of high power lasers has stimulated   a growing  interest towards   the experimental 
observation of    photon-photon scattering processes~\cite{Heinzl2006, Schlenvoigt2016, King} and electron positron 
pair creation~\cite{BuTu70}. 
In addition it  has 
provided  strong motivation for  their   theoretical study
in processes such as the scattering of  a laser pulse by a laser pulse
 \cite{Mourou, Marklund, DTom, Pare, DiPizzaReview, BattRizz, Koga, KarbsteinShai1,
 KarbsteinShai2},  the scattering of  $XFEL$ emitted photons \cite{Inada}, and 
 the interaction   of relatively long-wavelength, high intensity, laser light pulses  with short-wavelength $X$-ray photons
 \cite{Shanghai100PW}. 

The process of  vacuum polarization can be described within the framework of  the approximation 
using the Euler-Heisenberg Lagrangian \cite{HeisenbergEuler, Schw51}. Although this approximation  
is valid in the limit of colliding photons with relatively low energy and  of  low amplitude  electromagnetic pulses, 
it allows one to extend consideration over the non-perturbative theory.
Its applicability requires   the colliding photon energy to be  below the electron rest-mass energy, ${\cal E}_{\gamma}=\hbar \omega<m_e c^2$, 
and the electric field of  the colliding electromagnetic waves to be  below the critical field given by Eq. (\ref{eq:ES}).
 When writing the condition for the validity of the long-wavelength approximation given above  it was  assumed that the frequencies of the colliding photons are 
equal. If the frequencies are different, say  $\omega$ and $\Omega$ with $\Omega\neq\omega$, the low-frequency approximation requires  that 
\begin{equation}
 \omega \, \Omega <m_e^2c^4/ \hbar^2.
\label{eq:omOm}
\end{equation}
%

In the limit of  electromagnetic fields with extremely large amplitudes  approaching the QED critical field  $E_S$,  the nonlinear modification of the vacuum refraction index via the 
polarization of virtual electron-positron pairs leads 
to the decrease of  the propagation  velocity of counter-propagating 
electromagnetic waves \cite{Bialynicka, DittrichGies, KKB} while, on the contrary,  co-propagating waves do not change their propagation 
velocity because   co-propagating photons do not interact,  see  e.g. Ref. \cite{Zee}.

The nonlinear properties of  the QED vacuum in the long-wavelength, low frequency limit can find  a counterpart  in    those of nonlinear  dispersionless media, keeping however in mind 
that in QED there is no preferred frame where the nonlinear medium is at rest.
  In a material nonlinear medium with a  refraction index that depends 
 on the electromagnetic field amplitude an electromagnetic wave can evolve into a configuration 
with singularities \cite{Shock1, LL-EDCM}. The evolution of a finite amplitude wave  is 
accompanied by the  steepening of its wave front,  by the 
formation of shock-like waves, 
i.e. it is characterized by  a  processes leading to gradient catastrophes \cite{Whitham}.
In the case of the quantum vacuum,  corresponding phenomena have been investigated in Refs. \cite{Borel, LutzkyToll,  Boehl} and \cite{KKB}.
The occurrence  of singularities in the Euler-Heisenberg electrodynamics has been noticed in Refs. \cite{LutzkyToll, Borel}, indicated in computer simulations presented in Ref. \cite{Boehl}, and thoroughly studied in Ref. \cite{KKB}. 
 
In the present  paper, we analyze the interaction of finite amplitude,  counter-propagating  electromagnetic (e.m.) waves in a one dimensional (1-D) configuration.  The interacting  waves are assumed to be linearly polarized and to have the same polarization direction.
In such a configuration the propagation directions of the two colliding plane waves are collinear, and this collinearity is preserved by Lorentz boosts along the propagation direction. However, the Euler-Heisenberg Lagrangian is invariant under the full Lorentz  group. This makes it possible to use  the solutions that will be derived   in the following sections to  construct solutions that describe the interaction of plane waves colliding  at an angle, e.g.,  by considering Lorentz boosts in the direction perpendicular to the direction of the   polarization vector of the two colliding waves.   This extension of the results presented below may be of interest  in an experimental setting.

 The  hodograph transformation \cite{CouFr}  is a useful tool in the study of nonlinear  waves as it allows us   to obtain  
 a linear system of second order partial differential equations  (PDEs) instead of a system of second order quasilinear 
 PDEs. In the case of the e.m. 1-D configuration under study, this   transformation makes  the electric and the 
 magnetic fields play the role of the independent  coordinates.
  The hodograph transform has been adopted for a {\it non-dispersive}  formulation   of the electromagnetic field 
  equations in a  nonliner material medium, see e.g. Refs.\cite{RCA,Fusco}. 

The analysis described in the following sections allows us   to  find  exact solutions describing the nonlinear 
interaction of electromagnetic waves in vacuum
both in  the space-time coordinates  and in the hodograph variables,  to   formulate  a  perturbative approach  
that, in the limit of monochromatic waves,  does not lead to secularities  and to derive the dispersion relation of e.m. waves propagating in vacuum in the presence of  steady and uniform,  strong e.m. fields.

This article is organized as follows. In Sec. \ref{sec: HE-L} the Euler-Heisenberg Lagrangian is recalled   
and in  Sec. \ref{CPEW}  it is  specialized to the case  of counter-propagating  e.m. waves in a 1-D configuration and 
the corresponding nonlinear wave equation is derived  using the so-called light cone coordinates.  As an illustration,  
higher order terms that depend on the sixth power of the e.m. fields are included in the Euler-Heisenberg Lagrangian 
but, for the sake of  algebraic simplicity,  the contribution of these terms is neglected in some of the formulae in the  
present text. In Sec. \ref{SC} the conservations that arise from the translational  and from the Lorentz invariance of  
the 1-D Euler-Heisenberg Lagrangian are presented.  In Sec. \ref{LANIW} the linear case of non interacting waves is 
briefly described and  in Sec. \ref{PS}  perturbative solutions are obtained in light  cone coordinates.
In Sec. \ref{Char}  the derivation of the characteristics of the nonlinear wave equations is outlined, while in Sec. 
\ref{LIS}  exact self-similar solutions are derived. In Sec. \ref{LB} the dispersion equation of e.m. waves propagating  
in vacuum perpendicularly to large, steady and uniform, e.m. fields is presented.
In Sec. \ref{sec:hodograph} the  hodograph transform of the equations of nonlinear electrodynamics in vacuum is 
derived (see also Appendix \ref{AppA}) and 
in  Sec. \ref{NLIQEDHO} it is applied to  the study the nonlinear interaction of electromagnetic waves in the QED  
vacuum. In Sec. \ref{SCH} symmetries and conservations are reformulated  in the hodograph framework, while in 
Sec. \ref{HTLL} the expression  of non-interacting waves in hodograph variables is discussed (see also Appendix 
\ref{AppB}). In Sec. \ref{HPS} perturbative solutions are derived and  in Sec. \ref{HLIS} an exact selfsimilar solution is  
obtained.  In Sec. \ref{SF} the reduction of the hodograph equations to standard form is derived. This reduction  
makes  possible the use  of well known  expansion techniques for the solution of  linear PDEs with constant 
coefficients.  
Finally in Sec.\ref{CD}   a synthesis of the main results obtained  is given,   while in the Appendices \ref{AppA}, 
\ref{AppB}, \ref{AppC} some  proofs  and extensions of the  results in the main  text  and some additional 
mathematical developments are illustrated, including the identification of the function that plays the role of the 
Lagrangian in the  hodograph variables.

\section{Equations of nonlinear vacuum electrodynamics}
\label{sec: HE-L} 

The Euler--Heisenberg  Lagrangian is given by
\begin{equation}
\mathcal{L}=\mathcal{L}_{0}+\mathcal{L}', \label{eq:Lagrangian}
\end{equation}
where 
\begin{equation}
\mathcal{L}_{0}=-\frac{1}{16\pi}F_{\mu \nu}F^{\mu \nu}
\end{equation}
 is the  Lagrangian in classical electrodynamics, $F_{\mu \nu}$ is the electromagnetic field tensor 
\begin{equation}
F_{\mu \nu}=\partial_{\mu} A_{\nu}-\partial_{\nu} A_{\mu},
\end{equation}
with $A_{\mu}$ being the 4-vector of the electromagnetic field
and $\mu=0,1,2,3$. Here and below a summation over repeating indices is assumed.

In the Euler--Heisenberg theory, the QED radiation corrections are described by $\mathcal{L}'$ on the right hand side of Eq.(\ref{eq:Lagrangian}),  
which can be written as \cite{BLP-QED}
\begin{equation}
\label{eq:HELagr}
\mathcal{L}^{\prime}=
-\frac{m^4}{8\pi^2}\int^{\infty}_0 \frac{\exp{(-\eta)}}{\eta^3}\left[-(\eta {\mathfrak a}\cot \eta  {\mathfrak a}) (\eta {\mathfrak b}\coth \eta  {\mathfrak b}) +1-\frac{\eta^2}{3}({\mathfrak a}^2-{\mathfrak b}^2)\right] d \eta.
\end{equation}
Here the invariants  $ {\mathfrak a}$ and $ {\mathfrak b}$ 
can be expressed in terms  the Poincar\'e invariants 
\begin{equation} {\mathfrak F}=F_{\mu \nu}F^{\mu \nu} \quad {\rm and} \quad  {\mathfrak G}=F_{\mu \nu}\tilde F^{\mu \nu}
\end{equation}
 as 
\begin{equation}
\label{eq:abinv}
{\mathfrak a}=\sqrt{\sqrt{{\mathfrak F}^2+{\mathfrak G}^2}+{\mathfrak F}} \quad {\rm and} \quad {\mathfrak b}
=\sqrt{\sqrt{{\mathfrak F}^2+{\mathfrak G}^2}-{\mathfrak F}},
\end{equation}
respectively, where dual tensor $\tilde F^{\mu \nu}=\varepsilon^{\mu \nu \rho \sigma}F_{\rho \sigma}$
contains $\varepsilon^{\mu \nu \rho \sigma}$ being the Levi-Civita symbol in four dimensions. Here and in the following text, we use the units $c=\hbar=1$, and the electromagnetic field is normalized on the QED critical field $E_{S}$.

As explained in Ref. \cite{BLP-QED} 
the  Euler--Heisenberg Lagrangian in the form given by Eq.(\ref{eq:HELagr}) 
  should be used for obtaining an asymptotic 
series over the invariant electric field
 $ {\mathfrak a}$ assuming its smallness. 

In the weak field approximation the Lagrangian $\mathcal{L}^{\prime}$ 
is given by (e.g. see \cite{HeHe})
\begin{equation}
\mathcal{L}'=\frac{\kappa}{4}\left[{\mathfrak F}^2 
+ \frac{7}{4} {\mathfrak G}^2 +\frac{90}{315} {\mathfrak F} \left(  {\mathfrak F}^2 +\frac{13}{16}{\mathfrak G}^2  \right) \right]+...
\label{eq:mathcalL}
\end{equation}
 with the constant 
$\kappa=(e^4/360 \pi^2) {m}^4$.
In the Lagrangian (\ref{eq:mathcalL}) the first two terms on the right hand side   and the last two correspond respectively to  four and to six photon interaction.

\subsection{Counter-propagating electromagnetic waves}\label{CPEW}

In the following we  consider the interaction of counter-propagating electromagnetic waves with the same  linear polarization,  in which case the invariant ${\mathfrak G}$ vanishes identically.
\,
Such a field configuration  can be described  in a transverse gauge by a  vector potential having a single  component, ${\bf A}=A {\bf e}_z$, with ${\bf e}_z$
the unit vector along the $z$ axis. In terms of the light cone coordinates (see  e.g. Ref.~\cite{LCC})
\begin{equation}
x_+=(x+t)/\sqrt{2},\quad
x_-=(x-t)/\sqrt{2},
\end{equation} 
the vector potential ${\bf A}$ can be written as %
\begin{equation}
\label{eq:Aa}
A=a(x_+,x_-).
\end{equation} 
In these variables  the Lagrangian 
(\ref{eq:Lagrangian}) takes the form 
\begin{equation}
\mathcal{L}=-\frac{1}{4\pi}\left[wu-\epsilon_2 (wu)^2-\epsilon_3 (wu)^3  \right] 
\label{eq:lightcone-Lagrangian}
\end{equation}
where the field variables $u$ and $w$ are defined by 
\begin{equation}\label{fields}
u=\partial_{x_-}a \quad {\rm and} \quad w=\partial_{x_+}a 
\end{equation} 
and are related to the electric field $E=-\partial_t A$  (along $z$) and to the magnetic field $B = - \partial_x A$ (along $y$) by
\begin{equation}\label{fields2}
\ w  = -(E+ B)/\sqrt{2}, \quad  u = (E-B)/\sqrt{2} \quad {\rm and} \quad uw = (B^2 - E^2) /2.
\end{equation}

The dimensionless parameters
$\epsilon_2$ and $\epsilon_3$ in Eq.~(\ref{eq:lightcone-Lagrangian}) are given by 
\begin{equation}
\epsilon_2=\frac{2 e^2}{45\pi} =\frac{2}{45 \pi} \alpha \quad {\rm and} \quad
\epsilon_3=\frac{32 e^2}{315\pi} =\frac{32}{315 \pi} \alpha,
\end{equation}
 where $\alpha=e^2/\hbar c\approx 1/137$ is the fine structure constant, 
i.e., $\epsilon_2\approx 10^{-4}$ and $\epsilon_3\approx 2\times 10^{-4}$, respectively.

The field equations can be found by varying the Lagrangian:
\begin{equation}
\partial_{x_-}(\partial_u \mathcal{L})+\partial_{x_+}(\partial_w \mathcal{L})=0.
\end{equation}
 
As a result, we obtain the system of equations (see also Appendix \ref{AppA})
\begin{equation}
\label{eq:lightcone-1}
\partial_{x_-} w=\partial_{x_+} u,
\end{equation}
%
%
\begin{equation}
\label{eq:lightcone-2}
\partial_{x_+}  [u ( 1 - 2 \epsilon_2 uw - 3 \epsilon_3 u^2w^2) ]\, + \,  \partial_{x_-} [w ( 1 - 2 \epsilon_2 uw - 3 \epsilon_3 u^2w^2) ] =0.
\end{equation}
The first of these equations, Eq.(\ref{eq:lightcone-1}), 
is simply a consequence of the symmetry of the second derivatives, $\partial_{{x_-}{x_+}}a=\partial_{{x_+}{x_-}}a$  and it expresses the vanishing of the 4-divergence of the dual tensor $\tilde F^{\mu \nu}$.
By rearranging terms  and by inserting  Eqs.(\ref{fields}, \ref{eq:lightcone-1}),  Eq.(\ref{eq:lightcone-2})  can  be rewritten in the form of a  second order, quasi-linear  partial differential equation for the potential $a(x_+,x_-)$:
\begin{equation}
\label{eq:lightcone-a}
[1-uw(4\epsilon_2+9\epsilon_3uw)]\partial_{{x_-}{x_+}} a  =w^2(\epsilon_2 +3\epsilon_3 u w)\partial_{{x_-}{x_-}} a 
+u^2(\epsilon_2+3\epsilon_3 u w)\partial_{{x_+}{x_+}} a,
\end{equation}
where $u(x_+,x_-)$  and  $w(x_+,x_-)$  are defined by   Eqs.(\ref{fields}).

\subsection{Symmetries and conservations} \label{SC}

The Lagrangian (\ref{eq:lightcone-Lagrangian}), and thus Eq.(\ref {eq:lightcone-a}), are   invariant under the discrete transformation $x_+ \leftrightarrow x_-$ that interchanges $u$ and $w$.\,\, 
The Lagrangian (\ref{eq:lightcone-Lagrangian}) is  also invariant under  translations  along $x$ and $t$ and under Lorentz boosts along $x$. In fact   the four-vector potential  component $a$ is transverse to the boost 
and the field product  $uv$ is proportional to the Lorentz invariant ${\mathfrak F}$.\,\, 
In terms of the light cone coordinates  the corresponding infinitesimal  transformations can be written with obvious notation as (see also Ref.~\cite{LCC})
\begin{align} \label {symm0}
& x_+ \to x_+ + \delta_+, \quad  x_- \to x_- + \delta_-,\nonumber \\   &  x_+ \to (1 - \beta) x_+  \quad  {\rm and}  \quad  x_- \to (1 + \beta) x_- \, ,
\end{align}
and  the product $x^+ x^-$ is invariant under Lorentz boosts along $x$. 
According to Noether's theorem these  continuous symmetries  imply  the local conservation of the electromagnetic energy momentum tensor and of the ``barycenter'' (center of  the energy-momentum distribution)
which in light cone coordinates takes the form
\begin{align} \label {symm1}
& \partial_{x_+} T_{ww} +   \partial_{x_-}  T_{uw} = 0,  \qquad  \partial_{x_+} T_{wu} +   \partial_{x_-}  T_{uu} = 0 \nonumber \\  & \partial_{x_+} ( T_{ww} \, x_+  -  T_{wu} \,  x_-) +  \partial_{x_-} ( T_{uw} \,  x_+  -  T_{uu}  \,  x_-)   = 0, \end{align}
where
\begin{equation}\label{EMT}
T_{ij} = \frac{ \partial {\cal L} }{\partial(\partial_i a)} \, (\partial_j a)  - \delta_{ij} {\cal L}, \qquad  i,j = \pm , \qquad {\rm  and} \quad T_{++} \equiv T_{ww}, \, \quad T_{+-} \equiv T_{wu}, \quad  {\rm etc.}
\end{equation}
Neglecting for simplicity the $\epsilon_3$ term, from  ${\cal L }= -( uw - \epsilon_2 u^2w^2)/4\pi $ we have
\begin{align} \nonumber \label{EMT1}
&T_{ww}  = T_{uu} = \epsilon_2 u^2 w^2/4\pi,\\ &  T_{wu}  = - u^2 ( 1 -2 \epsilon_2 u w)/4\pi, 
\quad T_{uw}  = - w^2 ( 1 -2 \epsilon_2 u w)/4\pi.
\end{align}
where the trace and the determinant are Lorentz invariants.

\subsection{Linear  approximation and non-interacting waves} \label{LANIW}

In linear approximation Eqs.(\ref{eq:lightcone-2}, \ref{eq:lightcone-a}) take  the form 
\begin{equation}
\label{eq:lightcone-lino}
\partial_{x_+} u= -\partial_{x_-} w, \qquad 
\partial_{{x_-}{x_+}} a  = 0,
\end{equation}
where  Eq.(\ref{eq:lightcone-a}) has reduced to the standard linear wave equation in the light cone coordinates.

The first of Eqs.~(\ref{eq:lightcone-lino}), together with Eq.~(\ref{eq:lightcone-1}),
leads to the general solution  $u=f(x_-)$ and   $w=g(x_+)$ with  $f$ and $g$ arbitrary functions that are determined by the  initial conditions. 
For these solutions  the vector potential $a(x_+,x_-)$  takes the factorized form $a(x_+,x_-)=  a_+(x_+) + a_-(x_-) $ with $u =\partial_{x_-} a $ and $w = \partial_{{x_+}} a$. These solutions   describe
noninteracting electromagnetic waves propagating towards positive and negative directions along the $x$ axis, respectively.

Equations (\ref{eq:lightcone-1}, \ref{eq:lightcone-2}) allow for  particular solutions for which either $u=0$ or $w=0$, in which case   $w$ (or $u$) is an arbitrary
  function depending on the light cone variable $x_+$ (or $x_-$). These solutions  describe  finite amplitude electromagnetic waves  propagating  along $x$ 
  from right to  left (from left to  right) with  propagation velocity equal to the speed of light in vacuum.
Their shape  does not change in time and 
  the electric and magnetic field components are equal   $E=B=-w/\sqrt{2}$ and $T_{wu}  = - w^2 /(4\pi)  = -E^2/(2\pi)$, or equal and opposite  $E=-B=u/\sqrt{2}$\, and $T_{uw}  = - u^2 /(4\pi)  =- E^2/(2\pi)$.

\subsection{Perturbative solutions}\label{PS}

In the case of small but finite field amplitudes $u$,   $w$
 we can solve Eqs.(\ref{eq:lightcone-1},~\ref{eq:lightcone-2})  (or equivalently  Eq.(\ref{eq:lightcone-a}))  perturbatively   by expanding in powers of the field amplitudes, seeking 
solutions of the form $ u(x_-,x_+) =  u_0(x_-) +   u_1(x_-,x_+)$,  \,  $ w(x_-,x_+) =  w_0(x_+) +   w_1(x_+,x_-)$ \, (or equivalently of the form $ a(x_+,x_-) =  a_{0-}(x_-) +   a_{0+}(x_+)  +a_1(x_-,x_+)$).
Keeping  only cubic terms in the fields we obtain
\begin{align}
\label{eq:lightcone-pert1}
& u_1(x_-,x_+)  = \epsilon_2 \,   u_0^2(x_-) \, w_0(x_+)    +  \epsilon_2\,   [ \partial_{x_-}u_0(x_-) ]\, \int^{x_+} d x'_+ w_0^2(x_+'), \nonumber \\
& w_1(x_+,x_-)  = \epsilon_2 \,   w_0^2(x_+) \, u_0(x_-)    +  \epsilon_2\,   [ \partial_{x_+}w_0(x_+) ]\, \int^{x_-} d x'_-u_0^2(x_-'), \end{align}
where the two integral terms give the net  effect of the interaction between two finite length counter-propagating waves after the end of the interaction.
 Corresponding  results can be obtained by integrating directly the wave equation  for $a_1({x_+}{x_-}) $  up to cubic terms
\begin{equation}
\label{eq:lightcone-a1}
\partial_{{x_-}{x_+}} a_1({x_+},{x_-})  = \epsilon_2 [(\partial _{x_+} a_{0+})^2 \partial_{{x_-}{x_-}} a_{0-}  + (\partial _{x_-} a_{0-})^2 \partial_{{x_+}{x_+}} a_{0+}].
\end{equation}

\subsubsection{Phase shift induced by the interaction with a localized pulse}
Taking as an example  a monochromatic wave $u_0(x_-)= U_0 \cos{k(x-t)}$ interacting with  a localized counter propagating pulse $w_0$, such that $w_0(x_+) = 0$ both for   $x_+ > L$ and  for  $x_+ <-L$,
we find 
\begin{align} \label{eq:lightcone-pert2}
& u(x_-,x_+ <-L )  = u_0(x_-) = U_0 \cos{k(x-t)} ,\qquad {\rm and} \\  & u(x_-,x_+ >L )  = u_0(x_-)   +  \epsilon_2\,   [ \partial_{x_-}u_0(x_-) ]\, \int_{-L}^{L} d x'_+ w_0^2(x_+')  \nonumber \\
& = U_0 \left[ \cos(k{x-t)} - k \epsilon_2\sin{k(x-t)}\int_{-L}^{L} d x'_+ w_0^2(x_+')\right]
\nonumber \end{align}
which, to the considered expansion order, corresponds to a phase shift \cite{shift}.\, 

 \subsubsection{Interaction between monochromatic waves and propagation velocity}\label{secular}
 
 In the case of two interacting monochromatic waves (independently of their relative  frequencies)  Eqs.(\ref{eq:lightcone-pert1})  would lead to a secular behavior:
in other words,  the quadratic terms in the integrands Eqs.(\ref{eq:lightcone-pert1}) do not satisfy  in general the integrability conditions.
In order to restore integrability, we may uplift  an $\epsilon_2$ term in the expansion of the vector potential  $a(x_+, x_-)$ and define  the zeroth order solution  as 
 \begin{equation} \label{secul1}
 {\bar a}_{+0} (x_+ + \epsilon_2 s_+ (x_+,x_-)),\quad      {\bar a}_{-0} (x_- +  \epsilon_2 s_- (x_+,x_-)). \end{equation}
To leading order we recover Eq.(\ref{eq:lightcone-lino}),  while  two counter-terms are added to 
Eq.(\ref{eq:lightcone-a1}) that is changed into 
 \begin{align} \label{secul2}
& \frac{\partial^2 a_1(x_+,x_-) } {\partial x_+\partial x_- }  =  
\epsilon_2 [(\partial _{x_+} {\bar a}_{0+})^2 \partial_{{x_-}{x_-}} {\bar a}_{0-}  + (\partial _{x_-} {\bar a}_{0-})^2 \partial_{{x_+}{x_+}} {\bar a}_{0+}].
\\
&
-\epsilon_2  \frac{\partial}{\partial x_+} \left[     (\partial _{x_+} {\bar a}_{0+}) \frac{\partial s_+(x_+,x_-)}{\partial x_-}   \right]  - \epsilon_2  \frac{\partial}{\partial x_-} \left[    (\partial _{x_-} {\bar a}_{0-}) \frac{\partial s_-(x_+,x_-)}{\partial x_+}   \right] .
   \nonumber
 \end{align}
Neglecting higher order terms in $\epsilon_2$ we have
    \begin{align} \label{secul2bis}
& \frac{\partial^2 a_1(x_+,x_-) } {\partial x_+\partial x_- }  =  \epsilon_2  \frac{\partial}{\partial x_+} \left[   (\partial _{x_+} {\bar a}_{0+})\left(   (\partial _{x_-} {\bar a}_{0-})^2   -    \frac{\partial s_+(x_+,x_-)}{\partial x_-}  \right)  \right]    \nonumber\\
&  + \epsilon_2  \frac{\partial}{\partial x_-} \left[   (\partial _{x_-} {\bar a}_{0-}) \left(   (\partial _{x_+} {\bar a}_{0+})^2   -    \frac{\partial s_-(x_+,x_-)}{\partial x_+}  \right)  \right]   \end{align}
where we take
  \begin{align}  \label{secul2ter} & s_+(x_+,x_-)  = \int ^{x_- }    {d x^\prime_- }  \, (\partial _{x^\prime_-} {\bar a}_{0-})^2  \sim  \int ^{x_-}    {d x^\prime_- }  \,  (\partial _{x^\prime_-} {a}_{0-})^2, \quad \rightarrow s_+(x_+,x_-) \sim   s_+(x_-) 
   \nonumber\\  & s_-(x_+,x_-)  = \int ^{x_+ } d {x^\prime_+ } \, ( \partial _{x^\prime_+} {\bar a}_{0+})^2  \sim  \int ^{x_+} {d x^\prime_+ }    \,  (\partial _{x^\prime_+} {a}_{0+})^2, \quad \rightarrow   s_-(x_+,x_-) \sim   s_-(x_+)   \end{align}
 and set without loss of generality $a_1= 0$.
  Then to first order in $\epsilon_2$ the renormalized solutions read
 \begin{equation} \label{ren} a(x_+, x_-) =  a_+ \left (x_+ + \epsilon_2  \int ^{x_-}   {d x^\prime_- }  \,  (\partial _{x^\prime_-} { a}_{0-})^2 \right) + a_- \left (x_- + \epsilon_2  \int ^{x_+}   {d x^\prime_+ }   \, (\partial^\prime _{x_+} { a}_{0+})^2 \right) . \end{equation}
 
 The integrals in the arguments  lead to  two  amplitude dependent, inhomogeneous, propagation velocities   with absolute values smaller than the speed of light \cite{Burton}
  \begin{equation} \label{ren1 }  
  v_- (x_+)  = 1 - \epsilon_2 (\partial _{x_+} { a}_{0+})^2 , \quad   
  v_+ (x_-)  = 1 - \epsilon_2 (\partial _{x_-} { a}_{0-})^2 ,  
  \end{equation}
 and,   for  localized pulses,   to a phase shift  at the end of the interaction in agreement with 
 Eq.~(\ref{eq:lightcone-pert2}).  {This  amplitude dependent slowing of the wave propagation velocity 
 may lead to self-lensing and wave collapse of two counter-propagating pulses \cite{Marklund, EPJD}}.
 

  \subsubsection{Perturbed light cone variables} 
  
  Referring to Eq.(\ref{ren}), we note that the variables 
  \begin{equation} \label{ren2}  X_+ =  x_+ + \epsilon_2  \int ^{x_-} {d x^\prime_- }   \,  (\partial _{x^\prime_-} { a}_{0-})^2, \quad   X_- = x_- + \epsilon_2  \int ^{x_+}  {d x^\prime_+ }   \, (\partial _{x^\prime_+} { a}_{0+})^2 \end{equation}
 are ``gauge invariant'' and transform properly under  1-D Lorentz transformations, see the second line in  Eqs.(\ref{symm0}).
Thus the condition  $X_+ \, X_- = 0 $
defines  a Lorentz invariant {\it perturbed light cone}.  It is interesting to notice  that the  causal cone of a wave  event  is ``shrunk''  by a counter-propagating wave.

\subsection{Full solutions}\label{Char}

The characteristics $ x_\pm = \xi_\pm(s)$ of Eq.(\ref{eq:lightcone-a}), neglecting for the sake of notational simplicity the $\epsilon_3$ term, are given 
 by the quadratic equation 
\begin{equation}\label{char5}     
\epsilon_2 u^2(s) \, \left(\frac{d\xi_+}{ds}\right)^2 +  \epsilon_2\,  w^2(s) \, 
 \left(\frac{d\xi_-}{ds}\right)^2  +\, [1 - 4 \epsilon_2 \, u(s)  w(s)]  \,  \left(\frac{d\xi_+}{ds}\right) \left(\frac{d\xi_-}{ds}\right)^2 = 0,
  \end{equation}
  and are used in Ref.\cite{LutzkyToll} in order to construct ``simple wave'' solutions of  Eq.(\ref{eq:lightcone-a})  and  to prove that it admits the formation of discontinuities.
  \medskip 
  
 In the following  instead  we will seek for selfsimilar (scale invariant) solutions of Eq.(\ref{eq:lightcone-a}) by reducing it to an ordinary nonlinear differential equation.
  
  \subsection{Lorentz invariant solutions}\label{LIS}
  
  We look for solutions of the form  $a(x_+, x_-) =a(\rho)$,  with $\rho \equiv x_+x_-$  i.e.  for solutions that are constant along  the Lorentz invariant curves $x_+x_- = const.$  
Then, from Eq.((\ref{eq:lightcone-a})) we obtain 
 \begin{equation} \label{self3}
\left[1 -  4 \epsilon_2 \rho   \left(\frac{d a} {d \rho  } \right)^2  \right] \,\frac{d} {d \rho  }  \left (\rho   \frac{d a} {d \rho  }  \right) =  2  \epsilon_2 \rho  ^2 \left(\frac{d a} {d \rho  } \right)^2 \, \frac{d^2 a}{d \rho  ^2},  \end{equation}
which can be rewritten as 
\begin{equation} \label{self3bis}
\frac{d} {d \rho  }  \left (\rho   \frac{d a} {d \rho  }  \right) =  2  \epsilon_2 \frac {d}{d\rho} \left[ \rho^2 \left( \frac{d a}{d \rho  } \right)^3\right]
\end{equation}
and  yields the algebraic equation 
\begin{equation} \label{self3bis1}
\ \frac{d a} {d \rho  }  - 2  \epsilon_2 \rho\left( \frac{da}{d \rho  } \right)^3 =  \frac{C_2}{\rho} .\end{equation}
In the limit $\epsilon_2 \to 0$ we obtain (with $C_1$, $C_2$ arbitrary constants) 
 \begin{equation} \label{self3bis2}
 a =C_1 + C_2 \ln |{\rho}| ,\quad  w = C_2/x_+ , \,\, u = C_2/x_-\end{equation}
In these  solutions  the electric and the magnetic fields ``cumulate''  at $ x = \pm t$ where their amplitude diverges. 
In this case a power expansion in $\epsilon_2$  cannot be used,  while the approach of Eq.(\ref{secul1}) gives 
 \begin{equation} \label{self3ter}
  a = C_1 +C_2 \ln|{\bar \rho}|, \qquad {\rm with} \quad {\bar \rho}  =  X_+ \, X_- \
 =  x_+ x_- + \epsilon_2  \, C_2^2  \, ( x_+ / x_{+0}   + x_- / x_{-0}  -2). \end{equation}
 which  amounts to an amplitude dependent  shift in  the cumulation  coordinates  with 
  \begin{align} \label{self3quater}
 & w = \frac{C_2 \,  (x_- +   \epsilon_2  \, C_2^2/ x_{+0} )}{x_+ x_- + \epsilon_2  \, C_2^2  \, ( x_+ / x_{+0}   + x_- / x_{-0}  -2) },  \nonumber \\
 & u =  \frac{C_2 \,  (x_+ +   \epsilon_2  \, C_2^2/ x_{-0} )}{x_+ x_- + \epsilon_2  \, C_2^2  \, ( x_+ / x_{+0}   + x_- / x_{-0}  -2) }. \end{align}
 These Lorentz invariant solutions represent a special case of solutions obtained 
 in  the hyperbolic  coordinates
   \begin{equation} 
   \label{self4}
\rho  = x_+ x_- ,  \qquad \psi =(1/2)  \ln{(x_+/x_-)}  
\end{equation}
that are briefly discussed in Appendix \ref{AppB}.
  
  \subsection{Waves  in finite amplitude,    uniform   electric  and  magnetic fields in vacuum  }\label{LB}

Let us set
\begin{equation}
\label{craz1}
a(x_+, x_-) =  W_0\, x_+ + U_0\,  x_- + {\tilde a} (x_+, x_-) 
\end{equation}
with $ W_0 ,\, U_0$ uniform background  fields and  assume the a finite amplitude   field ordering 
%
\begin{equation}
\label{craz1b} {\hat W}_0 = \epsilon_2^{1/2} W_0 \sim   {\hat U}_0 = \epsilon_2^{1/2}  U_0 \sim {\cal O}(1), \qquad W_0, U_0 \gg \partial _{x_+}  {\tilde a} (x_+, x_-) ,\, \partial _{x_-}  {\tilde a} (x_+, x_-) .
\end{equation}
Then Eq.(\ref{eq:lightcone-a}) (with $\epsilon_3 =0$ for the sake of simplicity) becomes 
\begin{equation}
\label{craz2}
(1- 4\, {\hat U}_0{\hat W} _0)\, \partial_{{x_-}{x_+}} {\tilde a}  \, = \, {\hat W_0}^2 \, \partial_{{x_-}{x_-}}{\tilde a} \, + \, 
{\hat U_0}^2 \,  \partial_{{x_+}{x_+}}{\tilde a} , 
\end{equation}
which  is hyperbolic, and thus describes waves,  for $(1- 4\, {\hat U}_0{\hat W} _0)^2> 4 U_0^2 W_0^2$, i.e. for $  {\hat U}_0{\hat W}_0 <1/6$ and for $ {\hat U}_0{\hat W}_0 >1/2$.
Taking for the sake of simplicity 
\begin{equation}\label{craz3}
 {\tilde a} = {\tilde a}_0 \exp{[i (k_+ x_+ + k_- x_-)]} = {\tilde a}_0 \exp{[i (k x-  \omega t)]} ,
\end{equation}
with $k = (k_+ + k_-)/\sqrt{2}$ and $\omega = -(k_+ - k_-)/\sqrt{2}$,  we obtain the dispersion  equation 
\begin{align}
\label{craz4}
&(1- 4\, {\hat U}_0{\hat W} _0)\, k_+ k_- = {\hat W_0}^2 \, k_- ^2 +  {\hat U_0}^2 \, k_+ ^2, \quad {\rm i.e.} \\
&(1 - 4{\hat U}_0{\hat W} _0 + {\hat W} ^2_0 + {\hat U}^2_0)\, \omega^2  = (1 - 4{\hat U}_0{\hat W} _0 -{\hat W} ^2_0 - {\hat U}^2_0) \,   k^2 - 2 ({\hat W} _0^2  - {\hat U} _0^2) \, \omega k,\nonumber
\end{align}
In the two interesting limits of a purely electric ($W_0 = - U_0, \, E =  \sqrt 2 W_0 $) and purely magnetic 
($W_0 =  U_0, \, B =  - \sqrt 2 W_0 $) background fields we obtain
\begin{align}
\label{craz5}
&\omega_e^2  = [(1 +\epsilon_2\,  E^2)/(1 + 3\epsilon_2 \, E^2)]  k_e^2 \nonumber  \\
& \omega_b^2  = [(1 - 3\epsilon_2 \, B^2)/(1 - \epsilon_2 \, B^2)]  k_b^2 , \qquad  {\rm for} \,\, \epsilon_2\,  B^2 < 1/3,
\end{align}
that correspond to phase velocities smaller than the speed of light in vacuum (see reviews~\cite{DittrichGies} and \cite{ERBER} and 
references therein).

\section{Hodograph transform of  the equations of nonlinear electrodynamics in  vacuum}
\label {sec:hodograph}

A system of quasilinear partial differential equations, i.e. a system linear with respect to  the highest order  terms in the 
partial derivatives 
$\partial_{x_-}$ and $\partial_{x_+}$
with coefficients nonlinearly dependent on variables $u$ and $w$, admits the hodograph transformation \cite{CouFr}.
Assuming that  both $u$ and $w$ are not constant, we  perform the hodograph transformation  by  treating  them as 
coordinates, i.e.  we consider $x_-$ and $x_+$ as functions of $u$ and $w$:
\begin{equation}
x_-=x_-(u,w) \quad {\rm and} \quad
x_+=x_+(u,w).
\end{equation}

To transform the  system of Eqs.(\ref{eq:lightcone-1}, \ref{eq:lightcone-2}) to the  new coordinates 
$u$ and $w$ we need to express  the partial derivatives with respect to $x_-$ and $x_+$ in terms of derivatives 
with respect to $u$ and $w$. For a function $\Upsilon (x_-,x_+)$, using the chain rule, we have 
\begin{equation}
\partial_u \Upsilon= \partial_{x_-} \Upsilon  \partial_u {x_-}+ \partial_{x_+} \Upsilon  \partial_u {x_+}, \qquad
%
\partial_w \Upsilon= \partial_{x_-} \Upsilon  \partial_w {x_-}+ \partial_{x_+} \Upsilon  \partial_w {x_+}.
\end{equation}
Solving this system of equations with respect to $\partial_{x_-} \Upsilon$ and $\partial_{x_+} \Upsilon$
we obtain 
\begin{equation}
 \partial_{x_-} \Upsilon =J^{-1}\left(\partial_u \Upsilon \partial_w {x_+}-\partial_w \Upsilon \partial_u {x_+}\right), \qquad
%
 \partial_{x_+} \Upsilon =J^{-1}\left(\partial_w \Upsilon \partial_u {x_-}-\partial_u \Upsilon \partial_w {x_-}\right).
\end{equation}
Here $J=(\partial_u {x_-}\partial_w {x_+}-\partial_w {x_-}\partial_u {x_+})$ 
is the Jacobian of the coordinate transformation, which is assumed not to vanish.
Taking $\Upsilon$ equal  either $u$ or $w$ we find 
\begin{equation}
 \partial_{x_-} u =J^{-1}\partial_w {x_+}, \qquad  \partial_{x_+} u =-J^{-1}\partial_w {x_-}, \qquad
%
 \partial_{x_-} w =-J^{-1}\partial_u {x_+}, \qquad  \partial_{x_+} w =J^{-1}\partial_u {x_+}. 
 \end{equation}

Substitution of these relationships to Eqs.(\ref{eq:lightcone-1},~\ref{eq:lightcone-2}) yields
\begin{equation}
\label{eq:hod-1/}
\partial_u {x_+}=\partial_w {x_-},
\end{equation}
\begin{equation}
\label{eq:hod-2/}
[1-uw(4\epsilon_2+9\epsilon_3uw)]\partial_w{x_-} =-w^2(\epsilon_2 +3\epsilon_3 u w)\partial_w{x_+} - u^2(\epsilon_2+3\epsilon_3 u w)\partial_u {x_+}.
\end{equation}
From the system  (\ref{eq:lightcone-1}) and (\ref{eq:lightcone-2}) with coefficients nonlinearly dependent on $u$ and $w$ we have obtained a system 
of linear equations for $x_-$ and $x_+$. Equations (\ref{eq:hod-1/},\, \ref{eq:hod-2/}) are the  hodograph transform of Eqs.~(\ref{eq:lightcone-1},\,\ref{eq:lightcone-2}).  As is well known the nonlinearity of the original system 
is shifted from the field equation to the coordinate transformation. 

\section{Nonlinear interaction of electromagnetic waves in QED vacuum}\label{NLIQEDHO}

Introducing a potential function  $\Phi(u,w)$ such that the functions $x_-$ and $x_+$ are given by 
\begin{equation}
\label{Phi} x_-= \partial_u \Phi,  \quad {\rm and} \quad x_+=\partial_w \Phi, \end{equation}
 we can write Eqs.~(\ref{eq:hod-1/},~\ref{eq:hod-2/}) in the form
\begin{equation}
\label{eq:hod-main1}
[1-uw(4\epsilon_2+9\epsilon_3uw)]\partial_{u w} \Phi =-w^2(\epsilon_2 +3\epsilon_3 u w)\partial_{w w}\Phi - u^2(\epsilon_2+3\epsilon_3 u w)\partial_{u u} \Phi.
\end{equation}
In Appendix \ref{AppA}  an equivalent derivation of Eq.~(\ref{eq:hod-main1})  involving the momenta of the  Lagrangian  ${\cal L}$  is  presented.
  It is  also shown that the function $\Phi(u,w)$ is related to the   Lagrangian function for the hodograph equations.
  
\subsection{Symmetries and conservations in the hodograph representation}\label{SCH}

When applying  the hodograph transformation $x_\pm = x_\pm (u,w)$   a  conservation equation  of the form
  \begin{equation} 
  \partial_{x_+}  {\cal A}_+(x_+,x_-) +   \partial_{x_-}   {\cal A}_-(x_+, x_-) = 0, 
  \label{poi0} 
\end{equation} 
 becomes (see Appendix \ref{AppA})
\begin{equation} 
\label{poi}  \{  {\cal A}_+(u,w), x_- \}_{u,w} =    \{  {\cal A}_-(u,w), x_+ \}_{u,w}, 
 \end{equation} 
where ${\cal A}_\pm(u,w) =  {\cal A}_\pm(x_+ (u,w), x_+ (u,w) ), 
$ and 
$$\{ X , Y \}_{u,w} =  (\partial X/\partial u)(\partial Y/\partial w) - (\partial Y/\partial u)(\partial X/\partial w), $$ 
denotes Poisson brackets with respect to $ u$ and $w$.\,\,
Introducing the potential $\Phi(u,w)$, Eq.(\ref{poi}) can be rewritten as
\begin{equation} 
\label{poi1}  
\{    {\cal A}_+(u,w), \partial_u \Phi \}_{u,w} =    \{  {\cal A}_-(u,w), \partial_w \Phi  \}_{u,w}.  \end{equation} 
Taking either $  {\cal A}_+(u,w)  = T_{ww}$ and $ {\cal A}_-(u,w)  =  T_{uw}$ or  $ A(u,w)  = T_{wu}$ and $B(u,w)   =  T_{uu}$ as given by the expression of the energy-momentum tensor in  Eqs.(\ref{EMT1})
we  recover Eq.(\ref{eq:hod-main1}), here  for the sake of simplicity we  have  set $\epsilon_3 =0$.   Finally we note that Eq.(\ref{poi1}) can be rewritten as a  conservation law   in $u$-$w$ space  as 
\begin{equation}
\label{eq:hod-main1x}
\partial_w [(\partial_u  {\cal A}_+)( \partial_{u} \Phi )-  {(\partial _u \cal A}_-)( \partial_{w} \Phi)] + \partial_u [ (\partial_w {\cal A}_- )(\partial_{w} \Phi) -  (\partial_w {\cal A}_+)(\partial_{u} \Phi)]  = 0.\end{equation}
The  conservation  equation obtained by inserting the components of the energy-momentum tensor  in  Eqs.(\ref{EMT1}) into Eq.(\ref{eq:hod-main1x}) is related to the invariance of  Eq.(\ref{eq:hod-main1})  under the transformation 
\begin{equation}
\label{eq:hod-main1xx}
\Phi(u,w) \rightarrow  \Phi(u,w)  + \delta_+\, w \, + \,  \delta_-\, u, 
\end{equation}
which is the hodograph counterpart  of the coordinate  translations in Eq.(\ref{symm0}).
A similar procedure shows that the hodograph counterpart of the  conservation of the ``barycenter''  that is  given  in Eq.(\ref{symm1})  and that arises from the Lorentz invariance, yields a conserved quantity  that is quadratic  in $\partial_{u} \Phi, \, \partial_{u} \Phi$, see later Eq.(\ref{fin3ter}).

\subsection{Hodograph transformation in the linear limit} \label{HTLL}

In the linear limit, $\epsilon_2, \epsilon_3 \to 0$,  Eq.(\ref{eq:hod-main1}) reduces to
\begin{equation} \label{inv1}  
\ \frac{\partial ^2 \Phi(u,w)}{\partial u \,\, \partial w} =0, \qquad  {\rm i.e.,} \quad \Phi(u,w) = {\cal U} (u) + {\cal W}(w).\end{equation}
Here ${\cal U} (u) $  and ${\cal W}(w)$ correspond to counter-propagating 
non-interacting electromagnetic waves with
\begin{equation} \label{inv1bis}   
x_- = \ \frac{\partial \Phi(w,u) }{\partial u} =  \frac{\partial  {\cal U}(u) }{\partial u}, \qquad x_+ = \frac{\partial \Phi(w,u) }{\partial w} =  \frac{\partial  {\cal W}(w) }{\partial w} .  
 \end{equation} 

 The choice that corresponds to counterpropagating monochromatic waves is 
\begin{align} \label{inv2} &  {\cal U}_{k_u}  (u)   = \, \int_0^u \, du^\prime  \left[ - \psi_u  +\arcsin{(u^\prime /{\mathfrak A}_{u})}\right]/k_u + const,  \nonumber \\  
& {\cal W}_{k_w} (w) = \int_0^w \, dw^\prime  \left[ - \psi_w  +\arcsin{(w^\prime /{\mathfrak A}_{w})}\right] /k_w  + const, 
\end{align}
where ${\mathfrak A}_{w, u}$  are amplitudes,  $k_{u,w}$ ``frequencies'', $\psi_{u, w} $ are phases  and 
$$ \int_0^y \, dy^\prime \, \arcsin{(y^\prime)} =   y \arcsin{y}  +(1- y^2)^{1/2} - 1.$$ 
The definition domain is limited by  $|u/{\mathfrak A}_{u}| \, , \, |w/{\mathfrak A}_{w}| \, \leq 1$.   
By properly extending the image  domain of the $\arcsin$ function,
Eqs.~(\ref{inv2})  can be inverted as
\begin{equation}
 \label{inv2bis}  
u(x_-)  =  {\mathfrak A}_{u} \sin{(k_u x_- + \psi_u)} , \quad w(x_+)  = {\mathfrak A}_{ w} \sin{(k_w x_+ + \psi_w)} ,\end{equation}  
where the expressions inside each  oscillation half-periods  have been joined smoothly so as to cross over the points where the Jacobian of the  hodograph transformation vanishes. By redefining the origin of $x_\pm$ we can set   $\psi_u =   \psi_w\, = 0$ in agreement with Eq.(\ref{eq:hod-main1xx}).

In Appendix \ref{AppC} the role of the nonlinearity in the inverse hodograph transformation in the case of the 
superposition of two co-propagating monochromatic solutions is illustrated.

\subsection{Perturbative hodograph solutions}\label{HPS}

In analogy to the perturbative approach in ($x_+$-$x_-$) space  we can search for solutions 
of Eq.~(\ref{eq:hod-main1}) in the form of power  series  
$ \Phi =\Phi_0+\Phi_1+..., $
where $\Phi_0$ satisfies Eq.(\ref{inv1}).
%
To the first order to small parameters $\epsilon_2$ and  $\epsilon_3$ we obtain
\begin{equation}
\label{eq:hod-1}
\partial_{u w} \Phi_1 =-w^2(\epsilon_2 +3\epsilon_3 u w)\partial_{w w}\, {\cal W}(w) - u^2(\epsilon_2+3\epsilon_3 u w)\partial_{u u}\, {\cal U}(u),
\end{equation}
which  yields
\begin{align}
\label{eq:sol-1}
&& \Phi_1 =-\epsilon_2 u \int^w (w^\prime)^2 \partial_{w^\prime w^\prime}\, {\cal W}(w^\prime) dw^\prime -\frac{3}{2}\epsilon_3 u^2 \int^w (w^\prime)^3 \partial_{w^\prime w^\prime}\, {\cal W}(w^\prime) dw^\prime
\nonumber \\
&&- \epsilon_2 w \int^u (u^\prime)^2 \partial_{u^\prime u^\prime}\, {\cal U}(u^\prime) du^\prime-\frac{3}{2}\epsilon_3 w^2 \int^u (u^\prime)^3 \partial_{u^\prime u^\prime}\, {\cal U}(u^\prime) du^\prime \, .
\end{align}
For the choice of ${\cal W}(w)$ and ${\cal U}(u)$ in Eq.(\ref{inv2}) we obtain (for $\epsilon_3 =0$)
\begin{equation}
\label{eq:sol-2}
\Phi_1 =- \epsilon_2  \frac{u\, {\mathfrak A}^2_{w}}{2 k_w}  {\cal P}\left(\frac{w}{{\mathfrak A}_{w}}\right) - \epsilon_2  \frac{w\, {\mathfrak A}^2_{u}}{2 k_u}  {\cal P}\left(\frac{u}{{\mathfrak A}_{u}}\right) + const,
 \end{equation}
where 
$
{\cal P}(y) =
 \arcsin{(y)}
 - y 
( 1 - y^2  )^{1/2}
$.   
Inserting the zero-order solutions given in Eq.(\ref{inv2bis}) into  Eq.(\ref{eq:sol-2}) and inverting the hodograph 
transformation we can obtain  explicit expressions for 
$u(x_+,x_-)$ and  $w(x_+,x_-)$.  However, as noted above  for the corresponding perturbative  solutions in Eqs.
(\ref{eq:lightcone-pert2},\,\ref{eq:lightcone-pert1}), these expressions
include a term that exhibits a secular dependence on  the $x_+$, $x_-$  coordinates.  A procedure analogous to the  
one adopted in Eq.(\ref{secul1}) can  be used to remove this secular behavior 
as sketched in Appendix \ref{AppC}.

\subsection{Lorentz invariant solutions}\label{HLIS}

%
Equation (\ref{eq:hod-main1}) admits  self-similar solution when the function $ \Phi $ depends on the  Lorentz invariant variable $\xi=uw$ only.  These solutions are the hodograph counterpart of the solutions described by  Eqs.(\ref{self3bis}, \ref{self3ter}, \ref{self3quater})  in $x_+$, $x_-$ space.
For the function $\Phi(\xi)$ we obtain
\begin{equation}
\label{eq:hod-4photon-xi}
(1-4 \epsilon_2\xi-9\epsilon_3 \xi^2)(\Phi^{\prime}+\xi \Phi^{{\prime} {\prime}}) =-(2\epsilon_2 \xi^2-6\epsilon_3 \xi^3)\Phi^{{\prime} {\prime}},
\end{equation}
where $\Phi^{\prime}=d \Phi/d\xi$. Introducing the function 
$ U(\xi)=\Phi^{\prime}$
 Eq.(\ref{eq:hod-4photon-xi}) reduces  to
\begin{equation}
\label{eq:hod-U}
U^{\prime}+\frac{1-4 \epsilon_2\xi-9\epsilon_3 \xi^2}{\xi(1-2 \epsilon_2\xi-3\epsilon_3 \xi^2)}U=0.
\end{equation}
Integration of this equation yields 
%
\begin{equation}
\label{eq:hod-Uuw}
U(u w)=\frac{C}{u w(1-2 \epsilon_2 u w-3\epsilon_3 u^2w^2)}.
\end{equation}
For coordinates $x_-=w U$ and $x_+=uU$ we have   
\begin{equation}
\label{eq:hod-xmpuw}
x_-=\frac{C}{u(1-2 \epsilon_2 u w-3 \epsilon_3 u^2 w^2)}
\quad
{\rm and}
\quad
x_+=\frac{C}{w(1-2 \epsilon_2 u w-3 \epsilon_3 u^2 w^2)},
\end{equation}
which  in the limit $\epsilon_2= \epsilon_3 = 0$ coincide with Eq.(\ref{self3bis}). They can be rewritten as 
 \begin{equation}
\label{eq:hod-x}
x=\frac{2 C\, B}{(E^2-B^2)[1 + \epsilon_2 (E^2-B^2) -3\epsilon_3(E^2-B^2)^2/4]}
\end{equation}
and
 \begin{equation}
\label{eq:hod-t}
t=\frac{2 C\, E}{(E^2-B^2)[1 + \epsilon_2 (E^2-B^2) -3\epsilon_3(E^2-B^2)^2/4]}
\end{equation}

The solution given by Eq.(\ref{eq:hod-Uuw}) describes two counter-propagating 
electromagnetic pulses with the electric and magnetic fields  ``cumulating''  at the light cone  $x^2-t^2=0$  where 
the electric and magnetic fields tend to infinity.  Note that the position where  the cumulation  
occurs can be shifted by exploiting the translational invariance of the Lagrangian (\ref{eq:Lagrangian}), i.e.by 
looking for solutions   (see Eq.(\ref{eq:hod-main1xx}))  of the form  
$ \Phi(\xi) + \delta_+\, w \, + \,  \delta_-\, u $.

For all these  solutions the  Poincar\'e invariant $ {\mathfrak F}=F_{\mu \nu}F^{\mu \nu}=uw$ 
does not vanish for finite $x$ and $t$.
In the case of solutions (\ref{eq:hod-x},\ref{eq:hod-t}) the 
 dependence  ${\mathfrak F}$ on $t$ and $x$  is given by 
\begin{equation}
\label{eq:hod-Finv}
x^2-t^2=\frac{-4 C^2}{ {\mathfrak F}(1 + \epsilon_2  {\mathfrak F}-3 \epsilon_3  {\mathfrak F}^2/4)^2}.\nonumber 
\end{equation}
In the vicinity of the lines given by condition $x^2-t^2=0$ in the $(x,t)$ plane the expression (\ref{eq:hod-Finv}) cannot be used 
because here the electromagnetic field amplitude exceeds the critical QED field $E_S$. 

 The  Lorentz invariant solutions  derived above represent a special case of solutions obtained by using the  
 hyperbolic  coordinates in hodograph space 
$ \xi  =uw $ and $ \varphi =(1/2)  \ln{(u/w)}$. These solutions 
are briefly discussed in Appendix \ref{AppB}.

\subsection{Standard form of the hodograph wave equation}\label{SF}

The second order linear hyperbolic  PDE given by Eq.(\ref{eq:hod-main1})
can be set in the standard form (see e.g., \cite{PDE})
\begin{equation} \frac{\partial^2 \Phi}{\partial \zeta \partial \theta} \,   +  \, (\rm{\, lower \, order\, terms} ) = 0, \end{equation}
 by an appropriate  redefinition of the independent variables $u$ and $w$. 
For the sake of simplicity  in the following  this transformation will be performed here up to linear 
terms in $\epsilon_2$ and for $\epsilon_3 =0$.
We define the new independent variables 
\begin{align} \label{stl2} \nonumber & \zeta = u (1 - \epsilon_2 uw),  \quad  \theta = w (1 - \epsilon_2 uw) ,
\\  & u = \zeta  (1 +  \,  \epsilon_2 \zeta  \theta),  \quad  w = \theta  (1 + \,   \epsilon_2 \zeta  \theta) ,
\end{align} 
and obtain (here and in the following only linear terms in $\epsilon_2$ will be retained)
  \begin{equation}  \label{newv} 
  \frac{\partial^2  \Phi }{\partial \zeta  \partial \theta}   = 2 \epsilon_2 \left( \zeta \frac{\partial \Phi }{\partial \zeta} + \theta \frac{\partial \Phi }{\partial \theta} \right ) \left(1 - 8 \epsilon_2 \zeta\theta\right)^{-1}  \, 
\sim  \,  2 \epsilon_2 \left( \zeta \frac{\partial \Phi }{\partial \zeta} + \theta \frac{\partial \Phi }{\partial \theta} \right ).\end{equation} 
Note that the field variables $ \sqrt2\zeta = (E-B)[1 - \epsilon_2(B^2 - E^2)]$ and $ \sqrt2\theta = - (E+B)[1 - \epsilon_2(B^2 - E^2)]$
are directly related to the  perturbed  light cone variables $X_+,  X_- $   defined in Eq.(\ref{ren2})
since
\begin{equation}  \label{stl2b}   \theta  = \frac{\partial a}{\partial X_+} \quad {\rm and} \quad  \zeta = \frac{\partial a}{\partial X_-}.\end{equation}
Setting now $\Phi(\zeta,\theta) = \Phi_o(\zeta,\theta) \, (1 + 2 \epsilon_2\zeta \theta) $ we obtain  (to first order) the constant coefficient hyperbolic PDE
 \begin{equation}  \label{plasma0} \frac{\partial^2  \Phi_o(\zeta,\theta)}{\partial \zeta \, \partial \theta}   = 2 \epsilon_2 \Phi_o(\zeta,\theta), \end{equation} 
which is isomorphic to the equation for  linear  transverse e.m. waves in a uniform plasma. 

The solutions of Eq.(\ref{plasma0}) can be written in the general superposition form 
\begin{equation} \label{Sol0}  \Phi_o (\zeta,\theta)
  = \frac{1}{2\pi} \int_{-\infty}^{+\infty}  \int_{-\infty}^{+\infty} dk_\zeta dk_\theta   \, \delta(k_\zeta k_\theta +2\epsilon_2) \, {\tilde \Phi}_o(k_\zeta,k_\theta) \exp{[+i(k_\zeta \zeta + k_\theta \theta)]} +  {\cal C}{\cal C}  , \end{equation}
where the condition  $\delta(k_\zeta k_\theta +2\epsilon_2) $  accounts for the ``dispersion'' in   Eq.(\ref{plasma0}) and  ${\cal C}{\cal C}$ denotes complex conjugate.  This dispersion in the hodograph equation can be traced back  to  the nonlinearity of the wave equation in ($x_+$- $x_-$) space.

\subsubsection{Conservation equation}

If we add the two equations that we derive  by multiplying  Eq.(\ref{plasma0}) by 
$\partial \Phi_o (\zeta,\theta)/{\partial \zeta}$ and by $\partial \Phi_o (\zeta,\theta)/{ \partial \theta}$ 
respectively,  we obtain the following
conservation equation  
\begin{equation} \label{fin3ter} 
\frac{\partial }{\partial \theta}\left[   \frac{1}{2}  \left(  \frac{\partial  \Phi_o }{\partial \zeta} \right)^2- 
\epsilon_2  \Phi_o ^2 \right] + 
 \frac{\partial }{\partial \zeta}\left[   \frac{1}{2}  \left(  \frac{\partial  \Phi_o }{\partial \theta } \right)^2- 
 \epsilon_2  \Phi_o ^2 \right] 
  = 0 , 
  \end{equation} 
which is quadratic in the function $\Phi_0 (\zeta,\theta)$, and is related to the Lorentz invariance of the Lagrangian ${\cal L}$, see remark below Eq.(\ref{eq:hod-main1xx}).

\section{Conclusions and discussions}
\label{CD}

We have discussed within the framework of the Euler-Heisenberg  Lagrangian the main features of  the interaction in 
the quantum vacuum  
of counterpropagating electromagnetic fields.  
We have constructed explicit solutions of the  nonlinear 
hyperbolic wave equation obtained from the Euler-Heisenberg  Lagrangian within non-perturbative approach.  
We have 
used  a combination of analytical methods, involving  the direct search for solutions in space-time  light cone 
coordinates and the use of the hodograph transformation.
With the use of this transformation  the role
 of  the dependent and of the independent variables is  interchanged and, in the  restricted one-dimensional geometry 
 considered here,  the wave equation turns out to be a  linear hyperbolic equation to which standard solution methods 
 can be applied.
 When applying  the hodograph transformation the nonlinearity of the Euler-Heisenberg Lagrangian shifts to the 
 transformation itself which may be algebraically involved. 
 In addition, in the case of oscillatory fields the  implementation of the hodograph transformation  
 requires the smooth joining of piece-wise contributions since the transformation is not globally invertible.
 
With these analytical methods  we have constructed perturbative solutions and have identified exact selfsimilar 
solutions.  The relationship  between the properties of the solutions in both approaches has been discussed,  with 
special attention to 
the different forms that  conserved quantities take. 
These conservations arise   from the   translational invariance and from the Lorentz invariance of the 
Euler-Heisenberg Lagrangian. 
 
 We have shown that, in accordance with previous results in the literature,  
the interaction of two counter propagating pulses leads asymptotically only to a cumulative  phase shift, a result that 
can be understood in terms of the energy and momentum conservation of massless particles  in a  head-on collision.
On the contrary, during the  interaction of two  counterpropagating waves,  the propagation velocity of each of them is 
reduced  by a term 
that depends quadratically on the amplitude on the opposite propagating wave.
The phase velocity of linear waves propagating in vacuum in the presence of large,  steady and uniform 
electromagnetic fields (orthogonal to the direction of propagation)  has been derived and shown to be smaller than 
the speed of light in vacuum,
again by a term that, to leading order, depends on the square of the amplitudes of the steady electromagnetic fields.

Finally we note that  the same  analytical methods can be used to  find solutions of the so called Born-Infeld Equation 
\cite{BI}, see Ref.\cite{Whitham}.

\bigskip

\section*{Acknowledgments}

We thank  Drs.~G. Korn and N. N. Rosanov for fruitful discussions.
Supported by the project High Field Initiative (CZ$.02.1.01/0.0/0.0/15\_003/0000449$) from European Regional Development Fund. F.P.  thanks  the ELI--Beamlines project for its hospitality in September 2018.

\begin{thebibliography}{99}

\bibitem{QFT1} M. E. Peskin and D. V. Schroeder, {\it An Introduction to Quantum Field Theory}
(Addison-Wesley, Reading, MA.1995).

\bibitem{BLP-QED} V. B. Berestetskii, E. M. Lifshitz, and L. P. Pitaevskii, {\it Quantum Electrodynamics} (Pergamon, New York, 1982).

\bibitem{QFT2} S. Weinberg, {\it The Quantum Theory of Fields: Volume 1} (Cambridge University
Press, Cambridge, 1995).

\bibitem{QFT3} S. Weinberg, The Quantum Theory of Fields: Volume 2 (Cambridge University
Press, Cambridge, 1996).

\bibitem{FJD52} F. J. Dyson, 
{\it Phys. Rev.} {\bf 85}, 631 (1952).

\bibitem{R70} V. I. Ritus, {\it Sov. Phys. JETP} {\bf 30}, 1181 (1970).

\bibitem{FS13} F. Strocchi,  {\it An Introduction to Non-Perturbative Foundations
of Quantum Field Theory} (Oxford University Press, New York, 2013).

\bibitem{VR02} V. Rubakov, {\it Classical Theory of Gauge Fields} (Princeton University Press, Princeton and Oxford, 2002).

\bibitem{Whitham}  G. B. Whitham, {\it  Linear and Nonlinear Waves} ( Wiley, New York, 1974).

\bibitem{NNR98} N. N. Rosanov, {\it JETP} {\bf 86}, 284 (1998).

\bibitem{SS00} M. Soljaci{\'c} and M. Segev, {\it Phys. Rev. A} {\bf 62}, 043817 (2000).

\bibitem{YuKPA03} Yu. S. Kivshar and P. Agrawal, {\it Optical Solitons. From Fibers to Photonic Cristals}, 
(Academic, New York, 2003).

\bibitem{SAU31} F. Sauter, 
{\it Zeit. f\"ur Phys.} {\bf 69}, 742 (1931).

\bibitem{HeisenbergEuler} W. Heisenberg and H. Euler, {\it Zeit. f\"ur Phys.} {\bf 98},  714 (1936).

\bibitem{GVD09} G. V. Dunne, {\it  Eur. Phys. J.} {\bf 55}, 327 (2009).

\bibitem{Baur} G. Baur, K. Hencken, D. Trautmann, S. Sadovsky, and Y. Kharlov, {\it Phys. Rep.} {\bf 364}, 359 (2002).

\bibitem{ATLASScattering} ATLAS Collaboration, {\it Nature Physics} {\bf 13},  852 (2017).

\bibitem{PLowdon} P. Lowdon, {\it Phys. Rev. D} {\bf 96}, 065013 (2017).

\bibitem{Inada} T. Inada, T. Yamazaki, T. Yamaji, Y. Seino, X. Fan, S. Kamioka, T. Namba, and S. Asai, {\it Appl. Sci.} {\bf 7}, 671 (2017).

\bibitem{Schw51} J. Schwinger, {\it Phys. Rev.} {\bf 82}, 664 (1951).

\bibitem{Borel} S. Chadha and P. Olesen, {\it Phys. Lett.} {\bf 72B}, 87 (1977).


\bibitem{King} B. King and T. Heintzl, {\it High Power Laser Science and Engineering} {\bf 4}, 1 (2016).

\bibitem{Heinzl2006} T. Heinzl, B. Liesfeld,  K.-U. Amthor, H. Schwoerer, R. Sauerbrey, and A. Wipf,
{\it Optics Express} {\bf 267}, 318 (2006).

\bibitem{Schlenvoigt2016} H.-P. Schlenvoigt, T. Heinzl, U. Schramm, T. E. Cowan, and R. Sauerbrey, 
{\it Phys. Scr.} {\bf 91},  023010  (2016).

\bibitem{BuTu70} F. V. Bunkin and I. I. Tugov, 
{\it Sov. Phys. Dokl.} {\bf 14}, 678 (1970).

\bibitem {Mourou} G. A. Mourou, T. Tajima, and S. V. Bulanov, {\it Rev. Mod. Phys.} {\bf 78},  309 (2006).

\bibitem{Marklund} M. Marklund and P. K. Shukla, {\it Rev. Mod. Phys.} {\bf 78}, 591 (2006).

\bibitem{DTom} D. Tommasini, A. Ferrando, and M. Seco, {\it Phys. Rev. A} {\bf 77}, 042101 (2008).

\bibitem{Pare}  A. Paredes, D. Novoa, and D. Tommasini,
{\it Phys. Rev. A} {\bf 90}, 063803 (2014).

\bibitem{DiPizzaReview}  A. Di Piazza, C. M\"uller, K. Z. Hatsagortsyan, and C. H. Keitel, {\it Rev. Mod. Phys.} {\bf 84}, 1177 (2012).

\bibitem{BattRizz}  R. Battesti and C. Rizzo, {\it Rep. Prog. Phys.} {\bf 76}, 016401 (2013).


\bibitem{Koga} J. K. Koga, S. V. Bulanov, T. Zh. Esirkepov, A. S. Pirozkhov, M. Kando, and N. N. Rosanov, {\it Phys. Rev. A} {\bf 86}, 053823 (2012).

\bibitem{KarbsteinShai1} F. Karbstein and R. Shaisultanov, {\it Phys. Rev. D} {\bf 91}, 113002 (2015).

\bibitem{KarbsteinShai2}  H. Gies, F. Karbstein, C. Kohlfuerst, and N. Seegert, {\it Phys. Rev. D} {\bf 97}, 076002 (2018).

\bibitem{Shanghai100PW} B. Shen, Z. Bu, J. Xu, T. Xu, L. Ji, R. Li, and Z. Xu, {\it Plasma Phys. and Contr. Fusion}
 {\bf 4}, 044002 (2018).

\bibitem{Bialynicka} Z. Bialynicka-Birula and I. Bialynicki-Birula, {\it Phys. Rev. D} {\bf 2}, 2341 (1970).

\bibitem{DittrichGies} W. Dittrich and H. Gies, {\it Probing the quantum vacuum. 
Perturbative effective action approach in quantum electrodynamics
and its application}, Springer Tracts Mod. Phys., {\bf 166}, 1 (2000).

\bibitem{KKB} H. Kadlecova, G. Korn, and S. V. Bulanov, 
{\it Phys. Rev. D} {\bf 99}, 036992 (2019).

\bibitem{Zee} A. Zee, {\it Quantum Field Theory in a Nutshell}, (Princeton University Press, 2010).


\bibitem{Shock1} A. V. Gaponov, L. A. Ostrovskii, and G. I. Friedman, {\it Izvestia VUZ Radiofizika} {\bf 10}, 1376 (1967).

\bibitem{LL-EDCM} L. D. Landau and E. M. Lifshitz, {\it Electrodynamics of Continuous Media} (Pergamon, Oxford, 1984).

\bibitem{LutzkyToll} M. Lutzky and J. S. Toll, {\it Phys. Rev.} {\bf 113}, 1649 (1959).

 \bibitem{Boehl}
P. Boehl, B. King, and H. Ruhl, 
{\it J. Plasma Phys.} {\bf 82}, 655820202  (2016).


\bibitem{CouFr} R. Courant and K. O. Friedrichs, {\it Supersonic Flow and Scock Waves} (Interscience, New York, 1948).

\bibitem{RCA} C. Rogers, H. I. Cekirge,  and A. Askar, 
{\it Acta Mechanica}, {\bf 26}, 59 
(1977).

\bibitem{Fusco}  D. Fusco, {\it Physics of the Earth and Planetary Interiors}, {\bf 50}, 46 (1988). 

\bibitem{HeHe} J. S. Heyl and L. Hernquist,
 {\it Phys. Rev. D} {\bf 55}, 2449 (1997).
 
 
\bibitem{LCC} Y. S. Kim and M. E. Noz, {\it Am. J. Phys.}  {\bf 50}, 721 (1982).


\bibitem{shift} A. Ferrando, H. Michinel,  M. Seco,   and D. Tommasini, {\it Phys. Rev. Lett.} {\bf 99}, 1150404  (2007). 

\bibitem{Burton}
S. P. Flood and D. A. Burton, {\it EPL}, {\bf  100},  60005 (2012).


\bibitem{EPJD} M. Marklund and J. Lundin, {\it Eur. Phys. J. D}, {\bf  55}, 319 (2009).


\bibitem{NOGO} 
A. D. Berm\'udez Manjarres and  M. Nowakowski, {\it  Phys. Rev. A}, {\bf 95}, 043820 (2017). 

\bibitem{PDE}   W. E. Williams, in {\it Partial Differential Equations}  (Oxford University Press, 1980).

\bibitem{ERBER} T. Erber, {\it Rev. Mod. Phys.} {\bf 38}, 4 (1966). 
 
 \bibitem{BI}  M. Born and L. Infeld, {\it Proc. Roy. Soc. A}, {\bf 144},   425 (1934).
 

\end{thebibliography}

\appendix
\chapter{}

\section{The hodograph transformation in differential form and Euler-Heisenberg momenta}
\label{AppA}

\subsection{Momenta of the Euler-Heisenberg Lagrangian}

We define the field momenta $\Pi_u$,   $\Pi_w$ in the standard way in terms of the  Lagrangian  ${\cal L}$  
\begin{equation} \label {stand0}
\Pi_u = \frac{\partial {\cal L}}{\partial( \partial a /\partial{x_-})} =  \frac{\partial {\cal L}}{\partial u} , \quad  \Pi_w = \frac{\partial {\cal L}}{\partial( \partial a /\partial{x_+})}  =   \frac{\partial {\cal L}}{\partial w} ,
\end{equation}
and  find
\begin{equation} \label{stand0bis}
\Pi_u  =  - \frac{w}{4\pi} ( 1 - 2\epsilon_2 uw - 3 \epsilon_3 u^2w^2), \quad  \Pi_w  =  - \frac{u}{4\pi}  ( 1 - 2 \epsilon_2 uw - 3 \epsilon_3 u^2w^2) .
\end{equation}
 The equations of motion take the form
 \begin{equation} \label{stand1}
\partial_{x_+} \Pi_w \, + \,  \partial_{x_-} \Pi_u =0 ,
\end{equation}
which leads to Eq.(\ref{eq:lightcone-2}) in the main text.

\subsection{Hodograph equations  in differential form}

Equations (\ref{eq:lightcone-1}, \ref{eq:lightcone-2}), or equivalently Eqs.(\ref{eq:lightcone-1}, \ref{stand1})  
can be written in the  2-form formalism  as 
 \begin{equation} \label{stand2b}
(\partial_{x_+} u)\, d x_+\wedge d x_-  - \,  (\partial_{x_-} w )\, d x_+\wedge d x_-   = d u \wedge d x_- +  d w \wedge d x_ + = 0,
\end{equation}
 \begin{equation} \label{stand2c}
(\partial_{x_+} \Pi_w )\, d x_+\wedge d x_-  + \,  (\partial_{x_-} \Pi_u)\, d x_+\wedge d x_-   = d \Pi_w \wedge d x_- -  d \Pi_w \wedge d x_ + = 0 .
\end{equation}

Taking $u$ and $w$  as independent variables in Eq.(\ref{stand2b})  (assuming that the  Jacobian of the transformation is different from zero) we obtain 
 \begin{equation} \label{stand3}
(\partial_w{x_-} )\, d u\wedge d w - \,  (\partial_u{x_+})\, d u \wedge d w   = 0, \quad \rightarrow  \quad  \partial_w{x_-} =  \partial_u{x_+},
\end{equation}
i.e. Eq.(\ref{eq:hod-1/}). Similarly, using   $\Pi_u, \Pi_w$ as  the independent variables  in Eq.(\ref{stand2c}) we obtain 
\begin{equation} \label{stand3b}
\frac{\partial x_+}{\partial \Pi_w} +   \frac{\partial x_-}{\partial \Pi_u}= 0 , \end{equation}
which leads to Eq.(\ref{eq:hod-2/}), after $\Pi_u,\Pi_w$ are expressed in terms of $u,w$ through Eq.(\ref{stand0bis}).
Conversely, we can express $u$, $w$ in terms of $\Pi_u,\Pi_w$  and write the whole system of the  hodograph equations in terms of the momenta $\Pi_u,\Pi_w$.

Note that the hodograph transformation procedure described above is  also applicable to  the more general case  with vector potential $A_z = A(x,y,t)$. In this case however it would  lead to nonlinear equations 
as can be easily seen e.g., by appropriately reformulating Eq.(\ref{stand2b})  as a 3-form ($ dx\wedge dt \, \rightarrow \, dx \wedge dy \wedge dt $).

\subsection{Conservations and Poisson Brackets}

We can rewrite the conservation equation  (\ref{poi}) in the differential form
\begin{align} 
& (\partial_{x_+}  {\cal A}_+ ) \, d x_+\wedge d x_- +   (\partial_{x_-}   {\cal A}_-) \, d x_+\wedge d x_- = 0 \nonumber \\
& \rightarrow \quad  d  {\cal A}_+ \wedge d x_- =    d  {\cal A}_- \wedge d x_+ 
\end{align} 
and, imposing the hodograph transformation,  we obtain
\begin{align} \label{poi1b}
& [(\partial_u  {\cal A}_+ )\, (\partial_w x_- ) - (\partial_u x_- )\, (\partial_w  {\cal A}_+ )] \, du \wedge d w  \nonumber \\ 
&
 = \, [(\partial_u  {\cal A}_- )\, (\partial_w x_+ ) - (\partial_u x_+ )\, (\partial_w  {\cal A}_-)] \, du \wedge d w  \nonumber \\
& \rightarrow  \qquad \{   {\cal A}_+(u,w), x_- \}_{u,w} =    \{   {\cal A}_-(u,w), x_+ \}_{u,w}. 
\end{align}

\subsection{Hodograph Lagrangian}

We  introduce an ``effective vector potential''  ${\cal V}(\Pi_+, \Pi_-)$ such that 
\begin{equation} \label{standL3}
\frac{\partial a} {\partial x_+} = \frac{\partial {\cal V} } {\partial\,  \Pi_w} , \qquad \frac{\partial a} {\partial x_-} = \frac{\partial {\cal V} } {\partial\,  \Pi_u}.
\end{equation} 
Expressing  $w$ and $u$ as functions of $\Pi_\pm$ 
from Eqs.(\ref{stand0bis}) with $\epsilon_3 = 0$, we obtain $w/u = \Pi_u/\Pi_w $ so that 
\begin{equation} \label{standL4}   
\Pi_u =  - \frac{w}{4\pi}  [ 1 - 2\epsilon_2 (\Pi_ w/\Pi_u) w^2], \quad  \Pi_w =  - \frac{u}{4\pi} [ 1 - 2 \epsilon_2  (\Pi_ u/\Pi_w)  u^2] .
\end{equation} 
A perturbative solution of these  cubic equations gives 
\begin{align} \label{standL4c}   
& w \sim  - 4\pi \, \Pi_u ( 1 + 32\, \epsilon_2 \Pi_ w\Pi_u) , \quad  u \sim  - 4\pi \, \Pi_w ( 1 + 32\, \epsilon_2 \Pi_ w\Pi_u),\nonumber \\
& {\cal V} \sim  - 4\pi \, \Pi_w\Pi_u ( 1 + 16\, \epsilon_2 \Pi_ w\Pi_u).
\end{align} 
Using Eq.(\ref{standL3})  we can rewrite Eq.(\ref{eq:hod-1/}) as
 \begin{equation} \label{standL5}
  x_+ = \frac{\partial \Phi}{\partial w}  =   \frac{\partial {\bar  \Phi}}{\partial_{\Pi_w} {\cal V}},\quad x_- = \frac{\partial \Phi}{\partial u}  =   \frac{\partial {\bar  \Phi}}{\partial_{\Pi_u} {\cal V}}, \quad {\rm with}  \quad  {\bar  \Phi}(\partial_{\Pi_{w}}{\cal V}, \partial_{\Pi_{u}}{\cal V}) =   {\Phi}(w,u).  \end{equation} 
Finally Eq.(\ref{stand3b}) becomes 
\begin{equation} \label{standL6}
\frac{\partial  }{\partial \Pi_w}  \frac{\partial {\bar  \Phi}}{\partial_{\Pi_w} {\cal V}}+   \frac{\partial }{\partial \Pi_u}  \frac{\partial {\bar  \Phi}}{\partial_{\Pi_u} {\cal V}} = 0 , \end{equation}
where the unknown  function ${\bar  \Phi}(\partial_{\Pi_{w}}{\cal V}, \partial_{\Pi_{u}}{\cal V})  $ plays the role of the Lagrangian for the equations in the hodograph variables (see also Eq.(\ref{eq:hod-main1})).

\chapter{}
\section{Hyperbolic coordinates}
\label{AppB}

Instead of $x_+$ and $x_-$ we  can use the   hyperbolic coordinates
 \begin{equation} \label{self4bis}
\rho = x_+ x_- = x^2 - t^2,  \qquad  \psi = (1/2)  \ln{(x_+/x_-)} = \frac{1}{2} \ln{\frac{1+ t/x}{1- t/x}} = \arctanh{(t/x)} .\end{equation}
Under an infinitesimal (finite) Lorentz transformation (see Eq.(\ref{symm0})) we have
\begin{equation} \label {symm00}
\rho  \to \rho,   \quad \psi \to \psi + \beta , \qquad  \quad  \left(\psi \to  \psi  +  \frac{1}{2} \ln{\frac{1+ \beta}{1- \beta}} = \psi  + \arctanh{(\beta)} \right)  .\end{equation}

\subsection{Lagrangian in  hyperbolic coordinates} 

Since the 
 Heisenberg  Lagrangian (\ref{eq:lightcone-Lagrangian}) is  Lorentz invariant, when expressed in  hyperbolic coordinates, it cannot depend explicitly on $\psi$.
Starting from the Action in $x_+, x_-$ variables, bringing it to $\rho, \psi$ variables ad using the fact that the Jacobian of the transformation is equal to one, the new Lagrangian (with $\epsilon_3 = 0$) reads:
 \begin{equation} \label{LL0}
(- 4\pi) {\cal L}_L(\rho,\psi) =   \rho   \left( \frac{\partial a}{\partial \rho }\right)^2 + \frac{1}{4 \rho } \left( \frac{\partial a}{\partial \psi}\right)^2 - \epsilon_2\left[   \rho   \left( \frac{\partial a}{\partial \rho }\right)^2 + \frac{1}{4 \rho } \left( \frac{\partial a}{\partial \psi}\right)^2        \right]^2 .
\end{equation}

The self-similar solution Eq.(\ref{self3})   corresponds to $\partial a/{\partial \psi} = 0$ and can be derived directly from the Lagrangian ${\cal L}_L(\rho,\psi) $  in the  convenient form given by Eq.(\ref{self3bis}).
In the linear limit $\epsilon_2 =0$ the  Lagrangian ${\cal L}_L(\rho,\psi) $  can be expanded into ``$\psi$-harmonics'' and leads to power-law solutions.  For $\epsilon_2\not= 0$ these harmonics are coupled.

\subsection{Hodograph equation  in  hyperbolic coordinates} 

In terms of  the variables  $\xi = u w$ and  $ \varphi =(1/2)  \ln{(u/w)}$  Eq.(\ref{eq:hod-main1}) (with $\epsilon_3 =0$) becomes
\begin{equation} \label{LL4}
 \left(1 - 4 \epsilon_2 \xi \right) \left[ \frac{\partial }{\partial \xi } \left(  \xi  \frac{\partial  \Phi }{\partial \xi } \right)- \frac{1}{4 \xi}  \frac{\partial^2  \Phi }{\partial \varphi ^2} \right] + \epsilon_2 \left( 2 \xi^2  \frac{\partial^2  \Phi}{\partial \xi^2} +  \frac{1}{ 2} \frac{\partial^2  \Phi}{\partial \varphi^2}\right) =0.
  \end{equation}
Since  Eq.(\ref{LL4}) is linear and  its coefficients are independent of $\varphi$, its solutions  be decomposed  into a two sided Poisson expansion i.e., in $\cosh{(\alpha \varphi)}$ and $\sinh{(\alpha \varphi)}$  terms   with $\alpha$ a real number. We obtain  a family of  ODEs that, with self evident notation, can be written as 
\begin{equation} \label{LL4b}
 \left(1 - 4 \epsilon_2 \xi \right) \left[ \frac{\partial }{\partial \xi } \left(  \xi  \frac{\partial  \Phi_\alpha }{\partial \xi } \right)- \frac{\alpha^2}{4 \xi}  \Phi_\alpha \right] + \epsilon_2 \left( 2 \xi^2  \frac{\partial^2  \Phi_\alpha}{\partial \xi^2} +  \frac{\alpha^2}{ 2} \Phi_\alpha \right) =0.
  \end{equation}
In the linear limit ($\epsilon_2 =0$)   the solutions of Eq.(\ref{LL4b}) are of the form $ \Phi = C_1 w^\alpha + C_2 u^\alpha$ and, for positive integer values of $\alpha$, can be used as a polynomial basis in the non-interaction limit.

\section{Nonlinear inversion of hodograph solutions}
\label{AppC}

As an illustration of the nonlinearity  that is intrinsic to the inversion of the hodograph transformation  we  consider  the superposition  of two  co-propagating waves (marked by the upper index 1 and 2) in the hodograph variables each of which would correspond separately to a monochromatic wave  in agreement with Eq.(\ref{inv2}):
\begin{align} \label{inv22}   
&{\cal U}_{k_u^{(1)} ,k_u^{(2)} }  (u)   = \, \int_0^u \, du^\prime  \left[ - \psi^{(1)}_u/k_u^{(1)} 
 +\arcsin{(u^\prime /{\mathfrak A}^{(1)}_{u})}/k_u^{(1)}  \right.  \nonumber\\  & \left. - \psi^{(2)}_u/k_u^{(2)} + \arcsin{(u^\prime /{\mathfrak A}^{(2)}_{u})}/k_u^{(2)} \right] + const . 
 \end{align} 
Using the identity 
$$  a \arcsin{y}  = - i \ln{\left[   i y  + (1 - y^2) ^{1/2} \right]^a},$$  
 after some algebraic  steps  we obtain from Eq.(\ref{inv1bis})
\begin{align} 
\label{inv222} 
& \exp{[i  (k_u^{(1)}  k_u^{2} )^{1/2} (x_- +  \psi^{(1)}_u/k_u^{(1)} + \psi^{(2)}_u/k_u^{(2)})]} \,  =  \, {\cal G}(u), 
\end{align}
where
$$
 {\cal G} (u)  =\left[   i u /{\mathfrak A}^{(1)}_{u}  +(1 - (u /{\mathfrak A}^{(1)}_{u})^2) ^{1/2} \right]^{(k_u^{(2)}/k_u^{(1)})^{1/2} }   \left[   i u /{\mathfrak A}^{(2)}_{u}  + (1 - (u /{\mathfrak A}^{(2)}_{u})^2) ^{1/2} \right]^{(k_u^{(1)}/k_u^{(2)})^{1/2} }
 $$ 
to be solved for $u = u(x_-) =  {\cal G}^{-1} (\exp{[i  (k_u^{(1)}  k_u^{2} )^{1/2} (x_- +  \psi^{(1)}_u/k_u^{(1)} + \psi^{(2)}_u/k_u^{(2)})]})$.

\subsection{Renormalized hodograph solutions for interacting waves}

In view of Eq.(\ref{secul1})  we can rewrite Eq.(\ref{inv2})  as 
\begin{align} \label{invR2} &  {\cal U}_{k_u}  (u- \epsilon_2  u^2 w, \epsilon_2 w)   = \, \int_0^{u- \epsilon_2  u^2 w} \, \frac{du^\prime/k_u}{ 1 - 2\epsilon_2 u^\prime w}  \left[\arcsin{\left( \frac{u^\prime  + \epsilon_2 u^{\prime\, 2 } w}{A_u}\right)
} - \right. \nonumber \\
&   \left. \epsilon_2 \, S_u \left(\arcsin{\left(\frac{u^\prime }{A_u}\right)} , \arcsin{\left(\frac{w}{A_w}\right)} \right ) \right]  + const. , \nonumber \\ 
 & {\cal W}_{k_w}  (w- \epsilon_2  w^2 u, \epsilon_2u )   = \, \int_0^{w- \epsilon_2  u w^2} \, \frac{dw^\prime/k_w}{ 1 - 2\epsilon_2 u w^\prime }  \left[\arcsin{\left( \frac{w^\prime  + \epsilon_2 w^{\prime\, 2 } u}{A_w}\right)
} - \right. \nonumber \\
&   \left. \epsilon_2 \, S_w \left(\arcsin{\left(\frac{u}{A_u}\right)} , \arcsin{\left(\frac{w^\prime }{A_w}\right)} \right ) \right]  + const.
 \end{align} 
 Equation (\ref{invR2})  can be inverted (to first order in $\epsilon_2$) as 
\begin{align} \label{invR2bis}  & u(x_-, \epsilon_2 x_+)  =  A_{u} \sin{(k_u x_- + \epsilon_2 \, S_u (\arcsin{(u /A_{u})} , \arcsin{(w /A_{w})})}  =  \nonumber \\ & = A_{u} \sin{(k_u x_- + \epsilon_2 \, S_u (k_u x_-, k_w x_+))} ,\nonumber \\  &
 w(x_+, \epsilon_2 x_-)  =  A_{w} \sin{(k_w x_+ + \epsilon_2 \, S_w (\arcsin{(u /A_{u})} , \arcsin{(w /A_{w})})}  =  \nonumber \\ & = A_{w} \sin{(k_w x_+ + \epsilon_2 \, S_w(k_u x_-, k_w x_+))} .
 \end{align}

\end{document}